# Experimental demonstration of continuous electronic structure tuning via strain in atomically thin MoS$_2$

Keliang He[1], Charles Poole[1], Kin Fai Mak[2,3], and Jie Shan[1*]

[1] *Department of Physics, Case Western Reserve University, 10900 Euclid Avenue, Cleveland, Ohio 44106, USA*
[2] *Laboratory for Atomic and Solid State Physics, Cornell University, Ithaca, New York 14853, USA*
[3] *Kavli Institute at Cornell for Nanoscale Science, Ithaca, New York 14853, USA*

[*]E-mail: jie.shan@case.edu

## Abstract

We demonstrate the continuous tuning of the electronic structure of atomically thin MoS$_2$ on flexible substrates by applying a uniaxial tensile strain. A redshift at a rate of ~70 meV per percent applied strain for direct gap transitions, and at a rate 1.6 times larger for indirect gap transitions, have been determined by absorption and photoluminescence spectroscopy. Our result, in excellent agreement with first principles calculations, demonstrates the potential of two-dimensional crystals for applications in flexible electronics and optoelectronics.

The ability to continuously tune a material's properties is one of the most unique features of two-dimensional (2D) crystals. Because of their atomic thickness, the electronic and optical properties of these materials are highly sensitive to external perturbations.[1-5] 2D atomic membranes are also known as 'ultrastrength materials' with high elasticity and Young's modulus.[1, 2, 4-10] They can even be strained to the intrinsic limit of ~ 25%, as recently demonstrated in both monolayer graphene[7] and $MoS_2$.[9, 10] The elastic strain field has thus been proposed as an effective approach for continuous band structure tuning in 2D crystals.[1, 2, 4, 5]

The strain effect on the electronic and vibrational properties of graphene has been extensively investigated.[1, 2, 11-16] Nonetheless, the high symmetry of the graphene honeycomb structure, makes band gap opening by strain difficult.[17-19] In contrast, atomically thin $MoS_2$, a semiconducting transition metal dichalcogenide (TMD) with lower symmetry, offers an excellent opportunity for band structure tuning via strain engineering. Monolayer $MoS_2$ consists of hexagonal planes of S and Mo atoms in a trigonal prismatic structure. Several distinctive electronic and optical properties including a crossover from an indirect gap to a direct gap in the limit of monolayer thickness,[20, 21] strong excitonic effects,[22-25] and the possibility of full optical control of the valley and spin occupation[26-31] have been recently demonstrated in this material. The observed changes in the nature of the band gap and its size as a function of layer thickness[20] are a direct consequence of the changes in orbital interactions along the *out-of-plane* direction. This result also suggests that strain can be similarly employed to tune the *in-plane* orbital interactions, thus affecting the electronic structures in 2D $MoS_2$. Furthermore, since the direct gap is only slightly lower in energy compared to the indirect gap in monolayer $MoS_2$,[24] relatively small tensile strains are expected to be adequate to cause a direct to an indirect gap transition (~ 2% uniaxial strain[32, 33]) and even a semiconductor-to-metal transition (~ 10% biaxial strain).[32-34]

A great number of first principles calculations have been recently reported on the effect of strain on the electronic, vibrational and chemical properties of 2D $MoS_2$ and other semiconducting TMDs.[32-42] Although the details and magnitude of the predicted effect depend on the level of approximations involved, these calculations all predict a redshift of the gap energy with tensile strain. This result can be qualitatively understood as a result of reduced orbital overlap and hybridization due to weakened atomic bonds. The calculations also show the possibility of reducing the effective carrier masses by tensile strain, which can be explored to improve the carrier mobility and the transport characteristics.[35, 37, 40, 42] While the strain effect on the vibrational properties of 2D semiconducting TMDs has been studied experimentally,[43] experimental study of the strain dependence of their electronic properties, however, remains unavailable.

In this letter, we report a systematic experimental study of the electronic structure as a function of strain in both mono- and bi-layer $MoS_2$. A uniaxial tensile strain was applied to the samples using a cantilever device, and the strain dependence of their electronic structure was investigated by optical absorption and photoluminescence (PL) spectroscopy. Under relatively small strains, a redshift of ~70 meV/% strain was observed for the direct-gap excitons for both mono- and bi-layer $MoS_2$, and a slightly larger (~1.6 times) redshift rate was observed for the indirect-gap transitions in bilayer $MoS_2$. These results are in excellent agreement with first principles calculations.[32-42] Furthermore, no dependence of the effect on strains applied along the zigzag and armchair crystallographic orientation was observed. This is consistent with the isotropic in-plane elasticity predicted by the threefold symmetry of $MoS_2$.[32, 36] Our investigation paves the way to mechanically engineer the electronic and optical properties of atomically thin $MoS_2$ and other TMDs, which have emerged as a new class of 2D semiconductors.

In our experiment, we used PMMA (ePlastics) as a flexible and transparent substrate to apply controllable and reproducible strains on the samples. Atomically thin $MoS_2$ samples of well-defined crystallographic orientation were deposited directly on PMMA by mechanical exfoliation[44] (Supporting Information S1) of bulk crystals (SPI). Regions of mono- and bi-layers were identified by their optical contrast (inset of Fig. 1a) and were further confirmed by absorption and PL spectroscopy[20] (Fig. 1b). Their crystallographic orientation was inferred from the rotational anisotropy of the second harmonic generation (SHG) process[45, 46] (Fig. 1c). Details on the determination of the sample thickness and crystallographic orientation will be described below.

Strain was applied to the samples through van der Waals coupling at the sample-substrate interface along either the armchair or zigzag direction of the crystal by bending the flexible substrate in two perpendicular directions using the cantilever setup (Fig. 1a).[15] The applied strain $\varepsilon$ is calibrated according to the deflection of the cantilever $\delta$ as $\varepsilon = \frac{3t\delta}{2L^2}\left(1-\frac{x}{L}\right)$ for small deflections.[47] Here $L$ is the length of the cantilever, $x$ is the distance from the sample to the clamped edge of the cantilever, and $t$ is the thickness of the substrate. To avoid sample slippage, we have kept the strain below 0.5%. And a thin layer of PMMA can also be deposited on top of the samples. We have also studied a large number of samples (Supporting Information S2) and assumed the above strain calibration to be accurate only for those samples, which showed large and repeatable strain effects.

The absorption spectrum of atomically thin $MoS_2$ samples was measured using broadband radiation from a super-continuum laser, which was focused onto the samples with a 50x microscope objective. The reflected light from the $MoS_2$ samples on the PMMA substrate ($R_{MoS_2+sub}$) and from the bare substrate ($R_{sub}$) was collected by the same objective and directed to a grating spectrometer equipped with a Charge Coupled Device (CCD). The sample absorbance $A$ at photon energy $\hbar\omega$ is then determined from the reflectance contrast $\frac{R_{MoS_2+sub} - R_{sub}}{R_{sub}} = \frac{4}{n_{sub}^2 - 1} A(\hbar\omega)$,[20] where $\frac{4}{n_{sub}^2 - 1}$ is a local field factor with $n_{sub}$ as the refractive index of PMMA.[48] PL was measured simultaneously using the same optical setup by exciting the samples with a single-mode solid-state laser centered at 532 nm. The laser power on the samples was kept below 1 and 50 µW, respectively, for absorption and PL measurements to avoid sample heating. And the typical data acquisition time is 5 and 20 seconds for absorption and PL measurements, respectively.

The absorption and PL spectrum of both mono- and bilayer MoS$_2$ under zero strain are shown in Fig. 1b. In the energy range of 1.55-2.30 eV two characteristic absorption features, known as the A (~1.90 eV) and B (~2.05 eV) excitons, were observed. They are associated with the direct gap transitions around the K point of the Brillouin zone. The absorbance of the bilayer sample is about 2 times of the absorbance of the monolayer sample outside the region of the exciton resonances, which serves as an unambiguous determination of the bilayer thickness.[20] In the PL spectra, often only the A exciton emission associated with the lowest energy direct gap transitions is seen for monolayer samples. In contrast, a characteristic low-energy emission feature (I) near 1.55 eV is observed for bilayer samples. This feature is associated with the indirect-gap transitions[20, 21] and is absent in monolayer MoS$_2$, which is a clear indication of a crossover from an indirect gap to a direct gap semiconductor in the limit of monolayer thickness. We note that there is a finite Stokes shift for the PL due to factors such as unintentional doping and shallow traps, which could also be strain dependent. As we show below, this effect is also sample dependent. We, therefore, rely primarily on absorption spectroscopy to determine the exciton energies. PL is used to determine the indirect gap transitions since their absorption is extremely weak.

The crystallographic orientation of monolayer $MoS_2$ samples was determined by second harmonic generation (SHG) rotational anisotropy[45, 46]. SHG is allowed in monolayer $MoS_2$ since it is non-centrosymmetric while the process is forbidden in bilayer $MoS_2$, since inversion symmetry is restored. In our experiment, linearly polarized optical pulses from a mode-locked Ti:Sapphire oscillator centered at 800 nm were used to excite monolayer $MoS_2$ samples under normal incidence. The reflected SH intensity polarized along the pump polarization direction was monitored while the samples were rotated about their surface normal. A typical angular dependence with a characteristic 6-fold pattern is illustrated in Fig. 1c, which arises from SHG in crystals with $D_{3h}$ symmetry. The symmetry of the atomic structure also predicts that the polarized SH intensity is maximized for excitation polarized along the armchair direction and zero along the zigzag direction. The crystallographic orientation of bilayer samples was inferred from the orientation of the attached monolayer regions.

We illustrate in Fig. 1d the absorption spectrum of a monolayer sample under 0.4% strain along the armchair and zigzag direction. Both the A and B exciton features exhibit a redshift compared to the case of zero strain. However, no difference is observed for the two distinct strain directions within experimental accuracy. This is consistent with the isotropic in-plane elasticity expected for a crystal with $D_{3h}$ symmetry. For small strains, the electronic structure has been shown to only respond to the 2D hydrostatic strain, independent of its direction.[32, 33, 36, 38] Below we present results only for strains applied in the zigzag direction.

The strain dependence of the absorption and PL spectra of monolayer $MoS_2$ is shown in Fig. 2a. The PL spectra are normalized to the A exciton peak. Both the A and B exciton absorption peaks redshift with increasing strain. Figure 2b summarizes the exciton absorption peak energies as a function of strain. For the relatively small strains investigated here, a linear dependence is observed. A redshift rate of 64±5 and 68±5 meV/% strain is extracted for the A and B exciton, respectively. Within experimental accuracy, the A-B exciton splitting remains ~146 meV for the entire range of applied strain in this work. The A exciton PL peak shows a similar strain dependence and a redshift rate compared to the A exciton absorption peak, which suggests that the Stokes shift in this sample is largely strain independent.

To understand qualitatively the experimental results, we note that although the excitonic effects are particularly strong in atomically thin $MoS_2$ due to reduced dielectric screening,[22-25] the exciton binding energies have been shown by first principles calculations to be almost strain independent.[36, 42] The measured strain dependence of the excitonic resonance energies can thus be related directly to the changes in the electronic band structure. The bands near the Fermi energy in $MoS_2$ are composed primarily of the Mo 4$d$ and S 3$p$ atomic orbitals, and the nature of the band gap and its size is determined by the $d$-manifold splitting due to the crystal field of the trigonal prismatic structure and Mo 4$d$ - S 3$p$ orbital hybridization. Application of tensile strains increases the Mo-Mo and Mo-S bond length (and decreases the S-S layer distance due to Poisson contraction).[32] The net effect is a reduction in the orbital hybridization and $d$-bandwidth, which is revealed as the reduced band gap and red shifted exciton resonance energies. The observed magnitude of the effect (~70 meV/% uniaxial strain) is also in good agreement with *ab initio* calculations (40-100 meV/%).[32-42] Furthermore, the A-B exciton splitting in monolayer $MoS_2$ corresponds mainly to the spin-orbit splitting of the lowest energy valence bands. The insensitivity of the latter to small strains is expected since the spin-orbit interactions arise from the inner parts of the atoms and are insensitive to the atomic bond lengths.

For comparison, we also investigate the strain effects on the electronic structure of bilayer $MoS_2$ (Fig. 3). Similar strain effects are observed for the A and B exciton energies with a redshift rate of 71±5 and 67±5 meV/% strain, respectively, from the absorption measurements (Fig. 3b). These values are comparable to that of monolayer $MoS_2$. The redshift rate of the A exciton energy extracted from the PL measurements (Fig. 3c), however, is smaller (48±5 meV/%). Such a discrepancy indicates that the Stokes shift of the A exciton PL is strain dependent, likely due to strain-induced defects. The overall PL intensity (not shown in the normalized PL spectra of Fig. 3a), as well as the relative strengths of the I and A feature, decreases with increasing strain, which is also consistent with the picture of strain-induced defects. Moreover, the PL feature I near 1.55eV, corresponding to indirect gap transitions, is observed to redshift with a larger rate of 77±5 meV/% strain than the A exciton PL energy (Fig. 3c). If the strain dependence of the Stokes shift is similar for both the A and I emission features, the strain effect is about 1.6 times larger for the indirect gap transitions than for the direct gap transitions. Such a trend arises from the difference in orbital composition and the corresponding orbital hybridization as a function of strain at different points of the Brillouin zone. We note that for an accurate determination of the indirect gap energy and its strain dependence a direct absorption measurement, such as photoconductivity spectroscopy,[20] is needed.

In summary, we have demonstrated that both the direct and indirect band gaps of atomically thin $MoS_2$ can be efficiently tuned by applying a uniaxial tensile strain. For both monolayer and bilayer samples, the A and B exciton resonance peaks, associated with direct gap transitions, redshift under tensile strains with similar rates, and a larger redshift rate is observed for the indirect band gap transitions in bilayer MoS$_2$. We also find that the uniaxial strain effect is independent of crystallographic orientation of the crystal within the range of 0.5% strain applied in this work. We note that the large strain effects observed in atomically thin MoS$_2$ have importance beyond this specific material. We anticipate that the elastic strain field can be used to tailor the electronic structure of other semiconducting transition metal dichalcogenides with properties similar to MoS$_2$. Also, a spatially inhomogeneous strain field can be engineered to generate spatially varying band structures in a homogenous material. Interesting band bending effects and mechanically induced photodiodes can be engineered, for instance, for solar energy harvesting.[36] The compatibility of 2D atomic crystals with flexible substrates and their mechanically tunable electronic and optical properties open up a new opportunity for applications of 2D crystals in flexible electronics and optoelectronics.

## Acknowledgements

We thank Drs. Zenghui Wang and Philip X.-L. Feng for helpful discussions. This work was supported by the National Science Foundation grant DMR-0907477 and the Research Corporation Scialog Program at Case Western Reserve University. Additional support for the optical instrumentation was provided by the National Science Foundation grant DMR-0349201.

## Supporting Information

Description on exfoliation of atomically thin MoS$_2$ samples on PMMA cantilevers with well-defined crystallographic orientations and histogram of the strain effects in all atomically thin

MoS$_2$ samples studied in this work. This material is available free of charge via the Internet at http://pubs.acs.org.

**Figure captions**

**Figure 1.** (a) Schematic diagram of a cantilever used in the experiment. It is made of a square shaped PMMA substrate of length $L$ = 9.5 cm and thickness $t$ = 0.245 cm. MoS$_2$ samples were deposited near a corner of the substrate. Strain was applied on the samples by clamping one edge (grey) and bending the opposite edge of the substrate. Strain on the sample was calibrated by the sample location $x$ and the cantilever deflection $\delta$ as described in the text. Insets: atomic structure of monolayer MoS$_2$ and optical image of one sample with both monolayer (1L) and bilayer (2L) regions. The scale bar is 5 $\mu$m. (b) Typical absorption (right axis) and PL (left axis) spectrum of mono- and bi-layer MoS$_2$. The absorbance of the bilayers is about twice of the monolayers. The PL spectra were first normalized according to the A exciton peak intensity of the monolayer sample and then vertically displaced. The bilayer PL was magnified by 5 times for clarity. The A and B resonance features correspond to excitons associated with the direct gap transitions around the K point of the Brillouin zone. The I emission feature corresponds to indirect gap transitions and is observable only in bilayer MoS$_2$. The sharp peak around 1.96 eV in the PL spectrum arises from Raman scattering of the excitation laser by the PMMA Raman mode at 2954 cm$^{-1}$. (c) Second harmonic intensity as a function of angle of rotation $\phi$ of a monolayer MoS$_2$ sample about its $z$-axis. Both the fundamental and SH radiation are linearly polarized at 0°. The 6-fold symmetry was first established with a coarse angular measurement. The dependence was then measured with a finer angular step in the first quadrant and extrapolated to the other three

quadrants. The solid line is a fit of the data to $c_1 + c_2 cos^2(3\phi)$ with fitting parameters $c_1$ and $c_2$. The maxima and minima of the SH signal correspond, respectively, to the light polarization along the armchair and zigzag direction. (d) Absorption spectrum of a monolayer MoS$_2$ sample for strain (0.4%) applied along the armchair and zigzag direction. The zero strain case is included for comparison and the spectra are vertically displaced for clarity.

**Figure 2.** (a) Absorption (left panel) and PL (right panel) spectrum of a monolayer MoS$_2$ sample under tensile strains up to 0.52% along the zigzag direction. The dashed blue and red lines are guide to the eye of the redshift of the peaks. The black dashed line at 1.96 eV is the PMMA Raman line at 2954 cm$^{-1}$, which is strain independent for the range of strains applied in this work. (b) Strain dependence of the A and B peak energies from the absorption measurement and the A peak energy from the PL measurement (symbols). Lines are linear fits to the data.

**Figure 3.** As in figure 2, but for a bilayer sample. Strain dependence of the A and B peak energies determined from absorption and of the A and I peak energies determined from PL is summarized in (b) and (c), respectively.

Figure 1

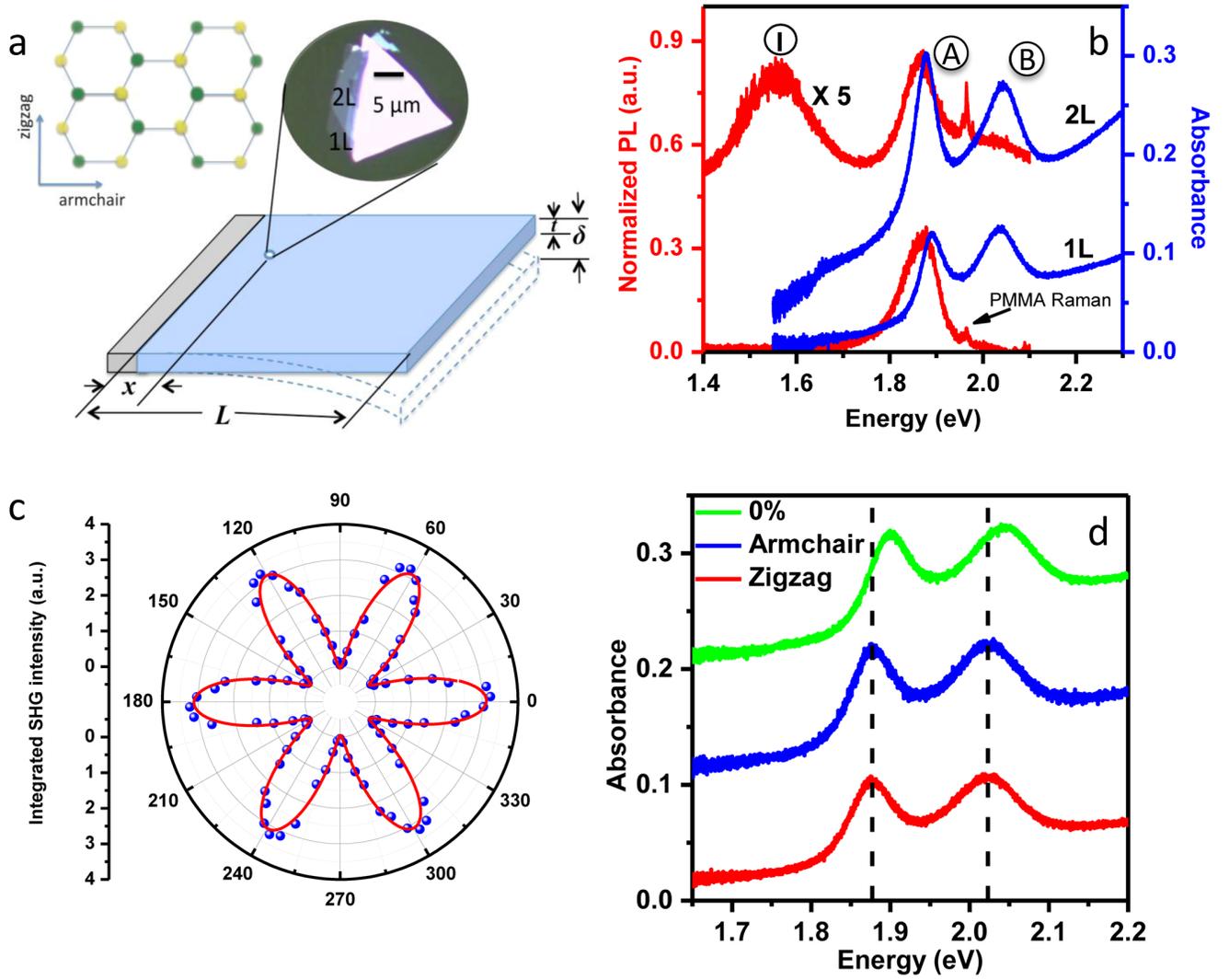

Figure 2

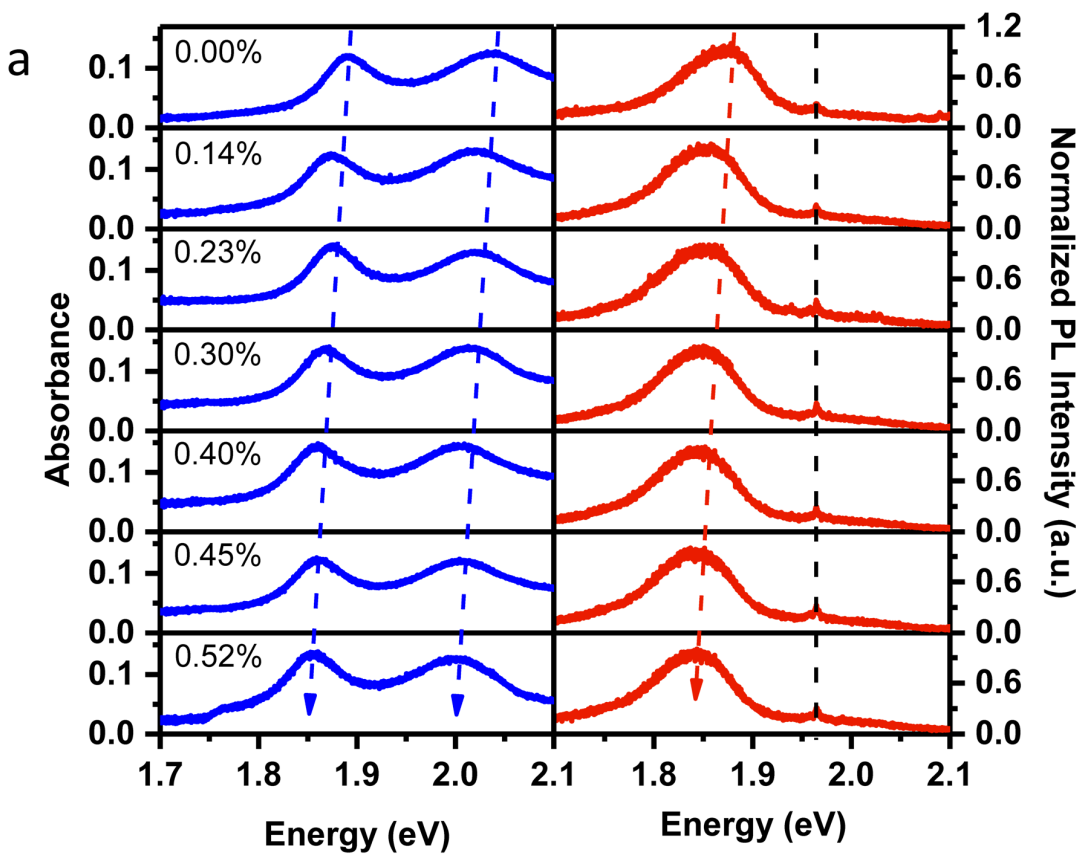

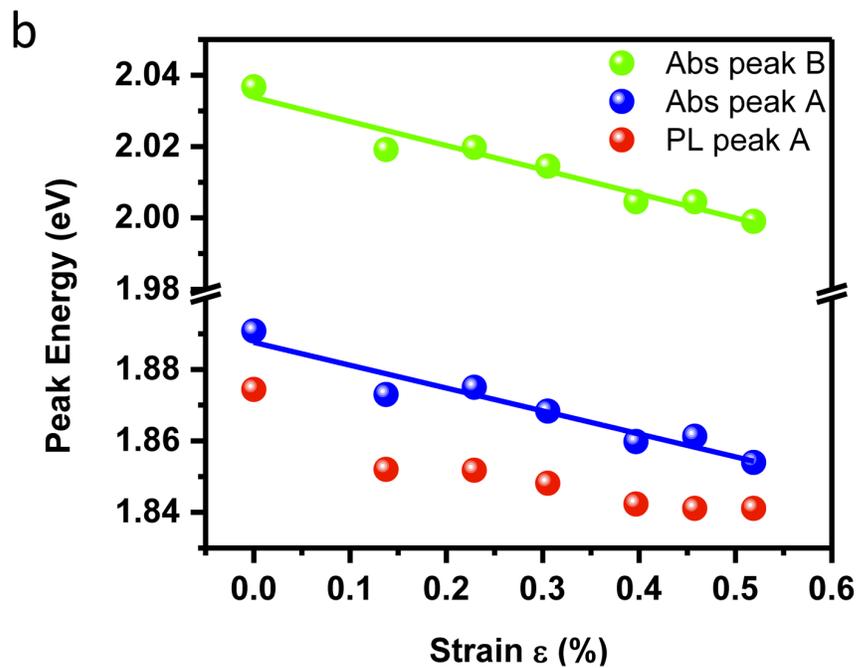

Figure 3

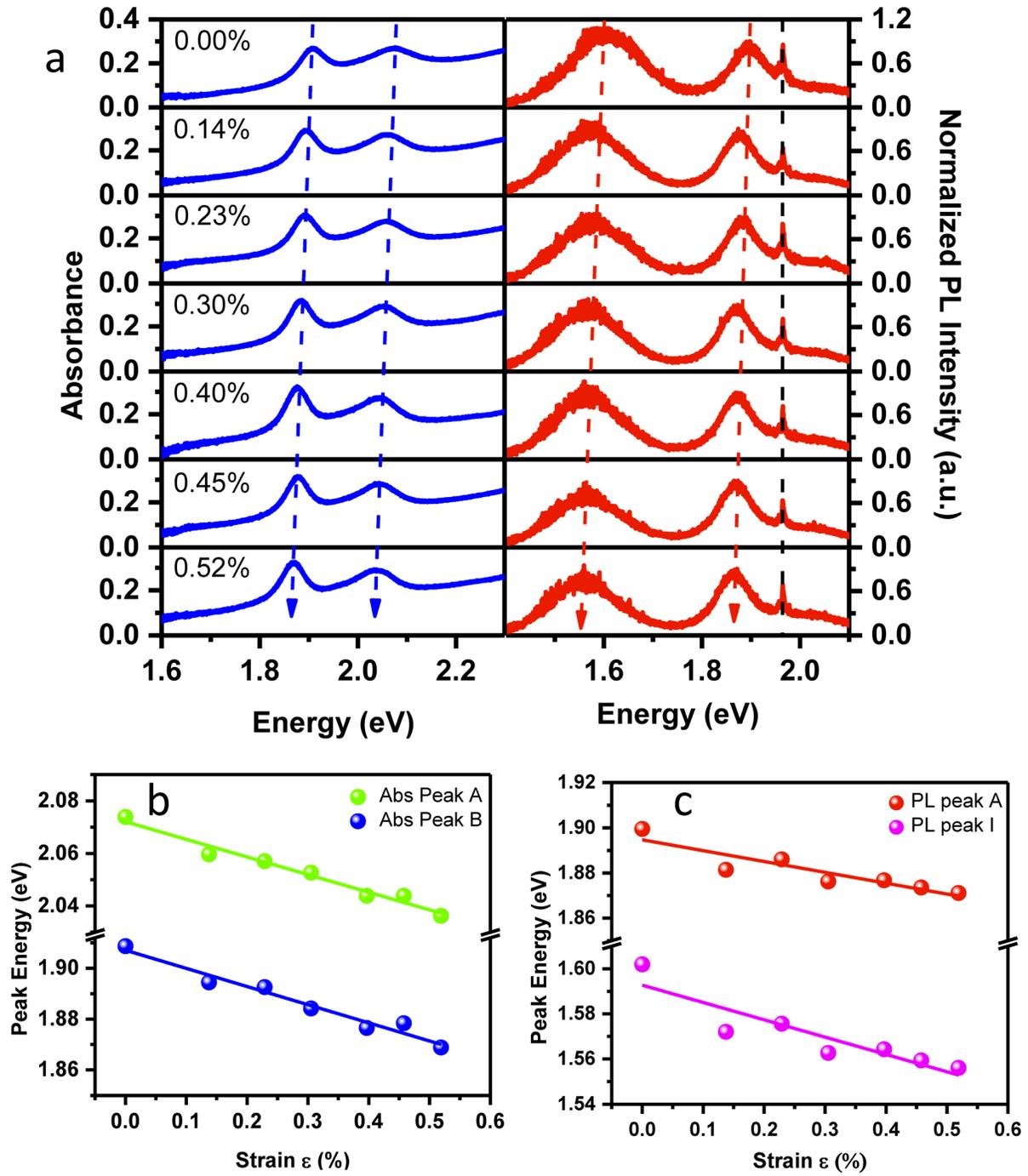

# Supplementary Information

# Experimental demonstration of continuous electronic structure tuning via strain in atomically thin MoS$_2$

Keliang He, Charles Poole, Kin Fai Mak, and Jie Shan

## 1. Exfoliation of atomically thin MoS$_2$ samples on PMMA cantilevers with well-defined crystallographic orientations

Square-shaped transparent PMMA substrates have been used as cantilevers to apply uniaxial tensile strain to the samples in two perpendicular directions (by clamping one side of the square and bending the opposite side). To maximize the applied strain on samples, atomically thin MoS$_2$ samples were deposited near a corner of the cantilever by mechanical exfoliation of bulk crystals (SPI). To align the crystallographic orientation of the samples with the substrate orientation, we rely on the fact that the samples cleaved from the same area of the bulk crystal have the same crystallographic orientation.

In the experiment, we used adhesive tape to repeatedly split bulk crystals into increasingly thinner layers following the standard mechanical exfoliation procedure.[S1] After the last step, two tapes with sample orientations as the mirror image of each other were obtained. One of the tapes was deposited onto a Si test substrate. Monolayer samples were identified and their crystallographic orientation was determined from the rotational anisotropy of the second harmonic generation (SHG) process as described in the main text. The other tape with inferred sample crystallographic orientation was then carefully aligned and deposited onto the PMMA substrate. Alignment of the zigzag (armchair) direction of the monolayer crystals with the substrate edges was verified by SHG to be within 5° for all the samples studied in this work.

## 2. Histogram of the strain effects in atomically thin MoS$_2$ samples

As described in the main text, strain was applied to the samples through van der Waals coupling at the sample-substrate interface using the cantilever setup. And the applied strain on the samples was calibrated according to the strain on the substrate surface. Such a calibration is not accurate if samples are wrinkled or pre-strained or there is sample slippage. In the experiment, we have kept the strain below 0.5% to avoid sample slippage. We have also studied a large number of samples and assumed the strain calibration to be accurate only for those samples, which show *large and repeatable* strain effects. A histogram of the strain effects for all the samples investigated in this work (23 monolayer samples and 5 bilayer samples attached to monolayer regions) is illustrated in Fig. S1. As an example we show the A exciton redshift rate determined from the absorption measurement and the calibrated strain.

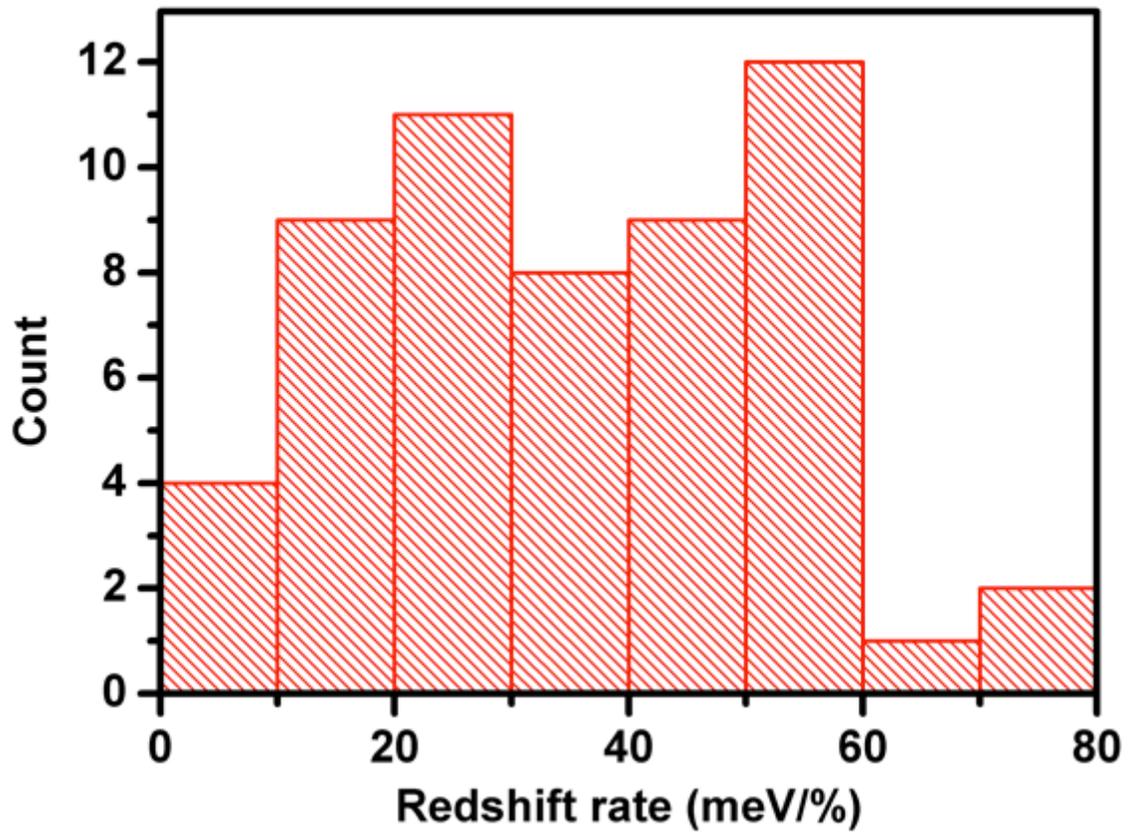

Figure S1. Histogram of strain effects in monolayer and bilayer $MoS_2$ samples studied in this work. The A exciton redshift rate was determined from the absorption measurement and the calibrated strain.

S1. Novoselov, K. S.; Jiang, D.; Schedin, F.; Booth, T. J.; Khotkevich, V. V.; Morozov, S. V.; Geim, A. K. *Proc Natl Acad Sci U S A* **2005,** 102, (30), 10451-10453.